 \documentclass[preprints,article,accept,moreauthors,pdftex]{mdpi}
\firstpage{1} 
\pubvolume{1}
\issuenum{1}
\articlenumber{0}
\pubyear{2021}
\copyrightyear{2021}
\datereceived{} 
\dateaccepted{} 
\datepublished{} 
\hreflink{https://doi.org/}

\usepackage{amssymb}
\usepackage{rotating}
\usepackage{tabulary}
\usepackage{tabularx,booktabs}
\usepackage[ruled,vlined]{algorithm2e}
\usepackage{algorithm2e}
\usepackage{xcolor}
\usepackage{soul,color}
\soulregister\cite7
\soulregister\ref7
\SetKw{KwBy}{by}

\Title{Spectroscopic Approach to Correction and Visualisation of Bright-Field Light Transmission Microscopy Biological Data}

\TitleCitation{Spectroscopic Approach to Correction and Visualisation of Bright-Field Light Transmission Microscopy Biological Data}

\Author{Ganna Platonova $^{1}$\orcidA{}, Dalibor \v{S}tys $^{1}$, Pavel Sou\v{c}ek  $^{1}$, Kirill Lonhus $^{1}$, Jan Valenta $^{2}$ and Renata Rycht\'{a}rikov\'{a} $^{1,}$*\orcidB{}}

\AuthorNames{Ganna Platonova, Dalibor \v{S}tys, Pavel Sou\v{c}ek, Kirill Lonhus, Jan Valenta and Renata Rycht\'{a}rikov\'{a}}

\AuthorCitation{Platonova, G.; \v{S}tys, D.; Sou\v{c}ek, P.; Lonhus, K.; Valenta, J.; Rycht\'{a}rikov\'{a}, R.}

\address{$^{1}$ \quad Institute of Complex Systems, Faculty of Fisheries and Protection of Waters, South Bohemian Research Center of Aquaculture and Biodiversity of Hydrocenoses, University of South Bohemia in \v{C}esk\'{e} Bud\v{e}jovice, Z\'{a}mek~136, 373 33 Nov\'{e} Hrady, Czech Republic;  
  gplatonova@frov.jcu.cz (G.P.);  stys@frov.jcu.cz (D.\v{S}.); psoucek@frov.jcu.cz (P.S.);  lonhus@frov.jcu.cz (K.L.) 
\\$^{2}$ \quad Faculty of Mathematics and Physics, Charles University, Ke Karlovu 3, 121 16 Prague, Czech Republic; jan.valenta@mff.cuni.cz  
}
\corres{Correspondence: rrychtarikova@frov.jcu.cz }

\abstract{The most realistic information about a transparent sample such as a live cell can be obtained   using bright-field light microscopy. Under high-intensity pulsing LED illumination, we captured a primary 12-bit-per-channel (bpc) response from an observed sample using a bright-field microscope equipped with a high-resolution (4872 $\times$ 3248) image sensor. In order to suppress data distortions originating from the light interactions with elements in the optical path, poor sensor reproduction (geometrical defects of the camera sensor and some peculiarities of sensor sensitivity), we propose a spectroscopic approach for the correction of these uncompressed 12 bpc data by simultaneous calibration of all parts of the experimental arrangement. Moreover, the final intensities of the corrected images are proportional to the photon fluxes detected by a camera sensor. It can be visualized in 8 bpc intensity depth after the Least Information Loss compression.}


\begin{document}


\section{Introduction}

Techniques of light microscopy give us a possibility to investigate features of various types of specimens and have been considered as the standard in observation of living cells  for decades~\cite{Ste03,Hel15,Mak15}. However, each of these microscopic techniques suffers from some disadvantages.

Unlike transmitted and reflected light microscopy techniques, fluorescence microscopy only allows the observation of specific structures which have been preliminarily labeled for fluorescence or excited with light for autofluorescence. Other limitations of   fluorescence microscopy are associated with photobleaching~\cite{Fad18} of applied dyes or autofluorophores; their phototoxicity influences biological processes in a live sample over time~\cite{Ich17,Lai17}. These aspects are avoided in contrast microscopy, e.g., phase contrast microscopy or differential interference contrast microscopy~\cite{Cre13}, where, naturally, some intracellular structures are enhanced, whereas some of them are suppressed by artificial light interferences~\cite{Cha15}. Nevertheless, for observation of the biological experiments using all microscopic techniques mentioned above, a grayscale digital camera is mostly sufficient~\cite{Bie17,Mig15,Dae16,Li15}, since shapes of the observed objects (e.g., cells or organelles) are mainly evaluated or, as in case of fluorescence microscopy, we observe a light of known wavelengths emitted from the sample.

The limitations of classical bright-field microscopy include low contrast for weakly diffracting, absorbing, and autofluorescent samples and low resolution due to the blurry appearance of out-of-focus material. However, bright-field transmission microscopy can allow the observation of unlabeled living cells and tissues and their internal structures and is of increasing interest due to recent experiments on super-resolution using video enhancement~\cite{Ryc17} and live cell dynamics~\cite{Ryc17b}. In label-free bright-field microscopy, the signal is generated mainly by diffraction of a wide range of light wavelengths in the sample. Light diffraction is a spectrally dependent phenomenon that can be utilized for determination of the transmission spectra of observed objects~\cite{Lon20}. Therefore, in order to describe the spectral properties of the observed sample, in the proposed microscope arrangement~\cite{Ryc17, Ryc17b}, the object response is captured by a color digital camera.

As a result of even slight defects of the camera sensor and aberrations in the optical path, images may suffer from strong distortions which disable  exact and simple data processing and analysis. Therefore, the microscope with the camera should be calibrated as any measurement device in order to achieve a more accurate representation of the object for further digital processing or visualization.

With the growing popularity of image processing automation, the question of technically correct and accurate images becomes even more acute. Different calibration approaches have been developed so far and are widely used in various microscopy techniques. The authors refer to the necessity of digital compensation of pixel-dependent noise specific to modern cameras used in fluorescent~\cite{Man19,Ber07} and single molecule localization microscopy~\cite{Bab19} as they affect the results of image analysis by software. In~\cite{Man20}, the authors introduced a theoretical model of noise sources and proposed their correction algorithm for fluorescence microscopy. On the other hand, there are approaches for pure data-driven determination of statistical properties of imaging sensors~\cite{Afa15}. In~\cite{Mic11}, the authors present a fast, data-driven method for correction of inhomogeneous fluorescence signals obtained via confocal microscopy when an image is multiplied by the inverse of the estimated spatially variable illumination gain. In~\cite{Mig19}, the authors discuss the impact of vignetting in application of light-field microscopy and introduce a calibration procedure to determine an optically correct reference point for microlens-based systems. Photometric calibration in spectral microscopy is discussed in~\cite{Thi10,You06,You08,Gru08}, where methods are proposed to compensate a non-uniform spectral response and quantum efficiency.

In the presented paper, we describe a method for capturing, correcting, and converting  12-bit raw images of unmodified biological samples obtained using a video-enhanced bright-field microscope. The image correction is based on the spectral characteristics of light captured by each camera sensor pixel during the calibration process. Advantages of the calibration method are demonstrated in both an image data visualization algorithm and simple image processing algorithms. The possible image visualization utilizes a full 8-bit-per-channel (bpc) range for transformation of corrected images (that depicted originally spatial distributions of spectral radiant fluxes). In this way, we acquired the most informative (in the sense of color and contrast) microscopic images of cells that can be mutually visually compared at the level of the whole image series. As we show further in this paper, the described method of the microscope calibration also brings  benefits to the subsequent microscopy image processing and analysis.

The calibration method may be technically improved but, from the conceptual point of view, is complete.
 
\section{Materials and Methods}

\subsection{Sample Preparation} \label{sample}

\subsubsection{Preparation of Live Cell Cultures}
We used two cell types for the experimental data collection: HeLa (human negroid cervix carcinoma, Sigma-Aldrich, Prague, Czech Republic, cat. No. 93021013) and L929 (mouse fibroblast, Sigma-Aldrich, Prague, Czech Republic, cat. No. 85011425). The cell lines were grown at low optical density overnight at 37 $^{\circ}$C, 5\% CO$_2$, and 90\% rH in a cell culture incubator. The nutrient solution for the HeLa cells consisted of 86.7\% minimal essential medium with high glucose (>1 g L$^{-1}$), 10\% fetal bovine serum (FBS), 1\% antibiotics and antimycotics, 1\% L-glutamine, 1\% non-essential amino acids, and 0.3\% gentamicin. The nutrient solution for the L929 cells consisted of 87.7\% Dulbecco's Modified Eagle Medium with high glucose (>1 g L$^{-1}$), 10\% FBS, 1\% antibiotics and antimycotics, 1\% L-glutamine, and 0.3\% gentamicin (all components purchased from Biowest Laboratories, Nuaill\'{e}, France).

For live cell imaging, fractions of live cell cultures were cultivated with relevant nutrient medium in a tissue culture dish   with a cover glass bottom (Ibidi, Gr{\"a}felfing, Germany, cat. No. 81158) covered by special glass lids (Ibidi, Gr{\"a}felfing, Germany, cat. No.~80050).

\subsubsection{Fluorescent Staining of HeLa Cells}

The fluorescent microscopy experiment was conducted with HeLa cells on tissue culture dishes (Ibidi, Gr{\"a}felfing, Germany). The nutrient medium was sucked out and the cells were rinsed by PBS. Then, the cells were treated by glutaraldehyde (3\%) for 5 min in order to fix cells in a gentle mode (without any substantial modifications in cell morphology) followed by washing in phosphate buffer (0.2 mol L$^{-1}$, pH 7.2) two times, always for 5~min. The cell fixation was finished by dewatering the sample in a concentration gradient of ethanol (50\%, 60\%, and 70\%) when each concentration was in contact with the sample for 5~min.

The dish with fixed HeLa cells was rinsed with distilled water, dried for approx. 1~h, then poured over the fluorescein solution (133 $\upmu$M) and washed on a minishaker at low rotational speed (approx. 50--100 rpm) for 30 min. After that, the dye solution was drawn off, the samples were slightly rinsed with distilled water (2$\times$) and finally dried again.

\subsection{High-Resolution Bright-Field Light Microscope} \label{microscope}

\begin{figure}[H]	 
\includegraphics[width=10 cm]{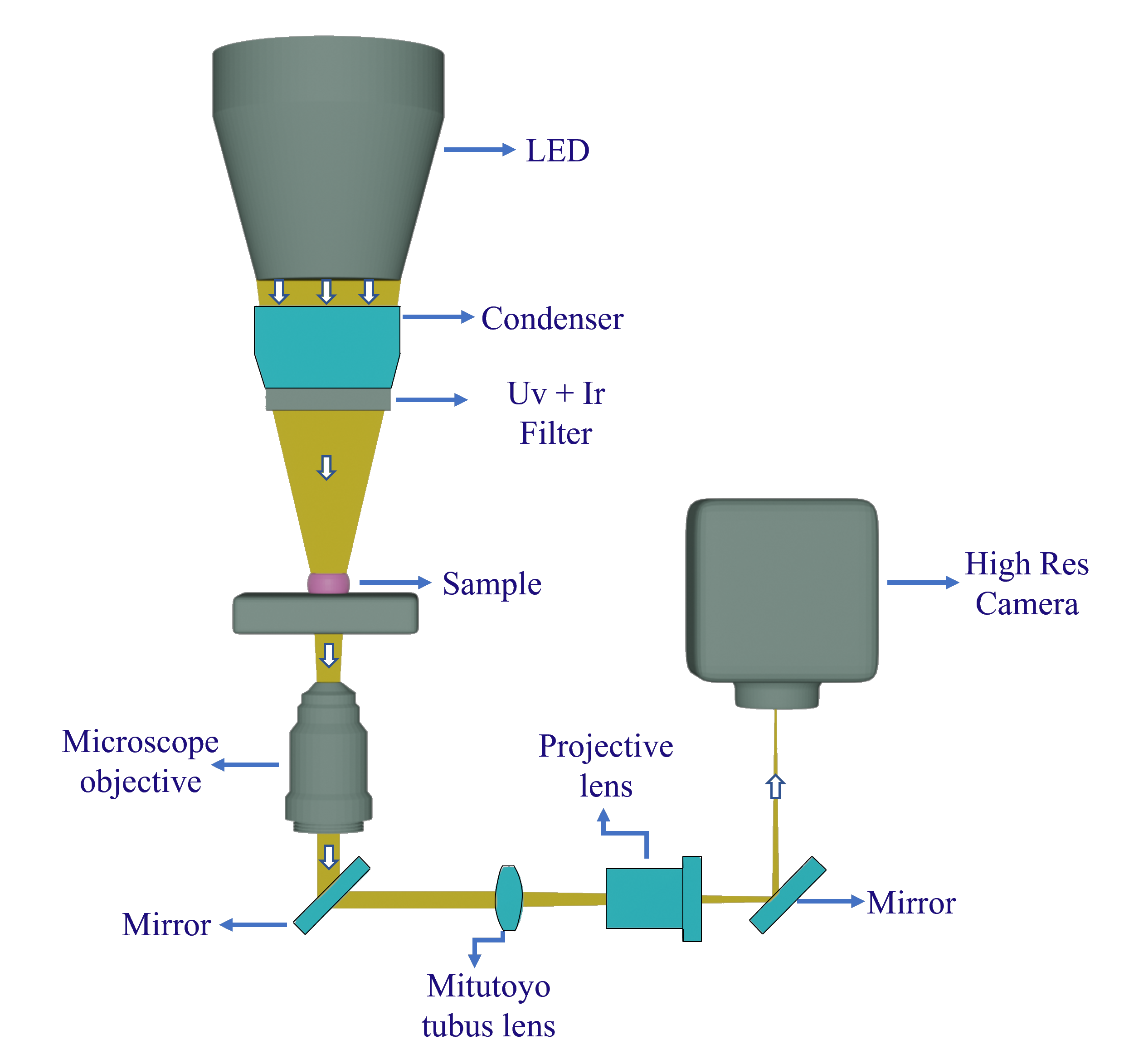}
\caption{The scheme of the nanoscope (ICS FFPW).  \label{Fig1}}
\end{figure}

In this paper, we present the process of calibration of the optical path and digital camera sensor of a high-resolution bright-field light transmission microscope, enabling observation of microscopic objects, further called the nanoscope~\cite{Ryc17,Ryc17b}. This microscope was developed by the Institute of Complex Systems (ICS) of the Faculty of Fisheries and Protection of Waters (FFPW) in collaboration with the Petr Tax--Optax company (Prague, Czech Republic). The scheme of the nanoscope components is shown in Figure \ref{Fig1}. The optical path consisted of two Luminus 360 light emitting diodes charged by the current up to 5000 mA. This allows the video enhancement~\cite{Lic96} by illumination of the sample by series (7$\times$) of light flashes with a period of 0.323 s and a duty cycle of 70\%; each of the series is followed by a light delay depending on the scanning frequency. In our research, two microscope objective lenses were used: 1. Nikon Plan 40$\times$/0.65, $\infty$/0.17, WD 0.56; 2. Nikon LWD 20$\times$/0.40, $\infty$/1.2, WD 3.1. A projective lens magnifies (2$\times$) and projects the image on a JAI camera with a 12 bpc color Kodak KAI-16000 image sensor of \mbox{4872 $\times$ 3248} image resolution. The process of capturing the primary signal was controlled by a custom control software. The optical system of the microscope was facilitated by infrared and ultraviolet filters (Edmund Optics, Barrington, NJ, USA).
 This prototype microscope allowed us to conduct experiments in time-lapse mode (a capturing of the image series at one focus position) and scanning of the interior of the cell along the optical axis in the z-scan mode (capturing in different z-positions). The z-scan can be performed automatically by   programmable mechanics with a step size down to 130.9 nm.

A time-lapse experiment (737 images) on mouse fibroblasts (see Section~\ref{sample}) was taken at the nanoscope set-up as follows: camera gain: 0, offset: 300, and exposure: 293.6 ms; LED current: 4500 mA; ILCX: $-106$ $\upmu$m; ILCY: $-201$ $\upmu$m; objective Nikon Plan: 40$\times$/0.65, $\infty$/1.17, WD: 0.56. A nanoscope image from a time-lapse experiment on HeLa cells taken at the set-up as follows: camera gain 0, offset 300, and exposure 295 ms; LED current 3500 mA; objective Nikon LWD 20$\times$/0.4, $\infty$/1.2, WD 3.1, closed NAMC2. 

In order to compare our custom microscope with conventional commercial microscopes, we used
\begin{enumerate}
\item A phase contrast microscope Nikon BioStation IMQ with a high-sensitivity 1.3 Mpx cooled monochrome camera, 40$\times$ objective magnification (Nikon, N.A. 0.8);
\item A bright-field microscope Olympus IX51, 40$\times$ objective magnification (objective LUC Plan FLN 40$\times$ 0.60 Ph2, $\infty$/0-2/NF 22),  camera Infinity 1 giving raw image data;
\item A fluorescent microscope Nikon Eclipse 80i, 40$\times$ objective magnification (Nikon Plan Fluor 40$\times$/0.75, Ph2 DLL, $\infty$/0.17, WD 0.66), mercury discharge tube, cell sample was stained by fluorescein.
\end{enumerate}
The comparative biological experiments were conducted on HeLa cells (see Section~\ref{sample}).

\subsection{Microscope System Calibration and Image Correction} \label{calibration}

\begin{figure}[H] 
\includegraphics[width=13 cm]{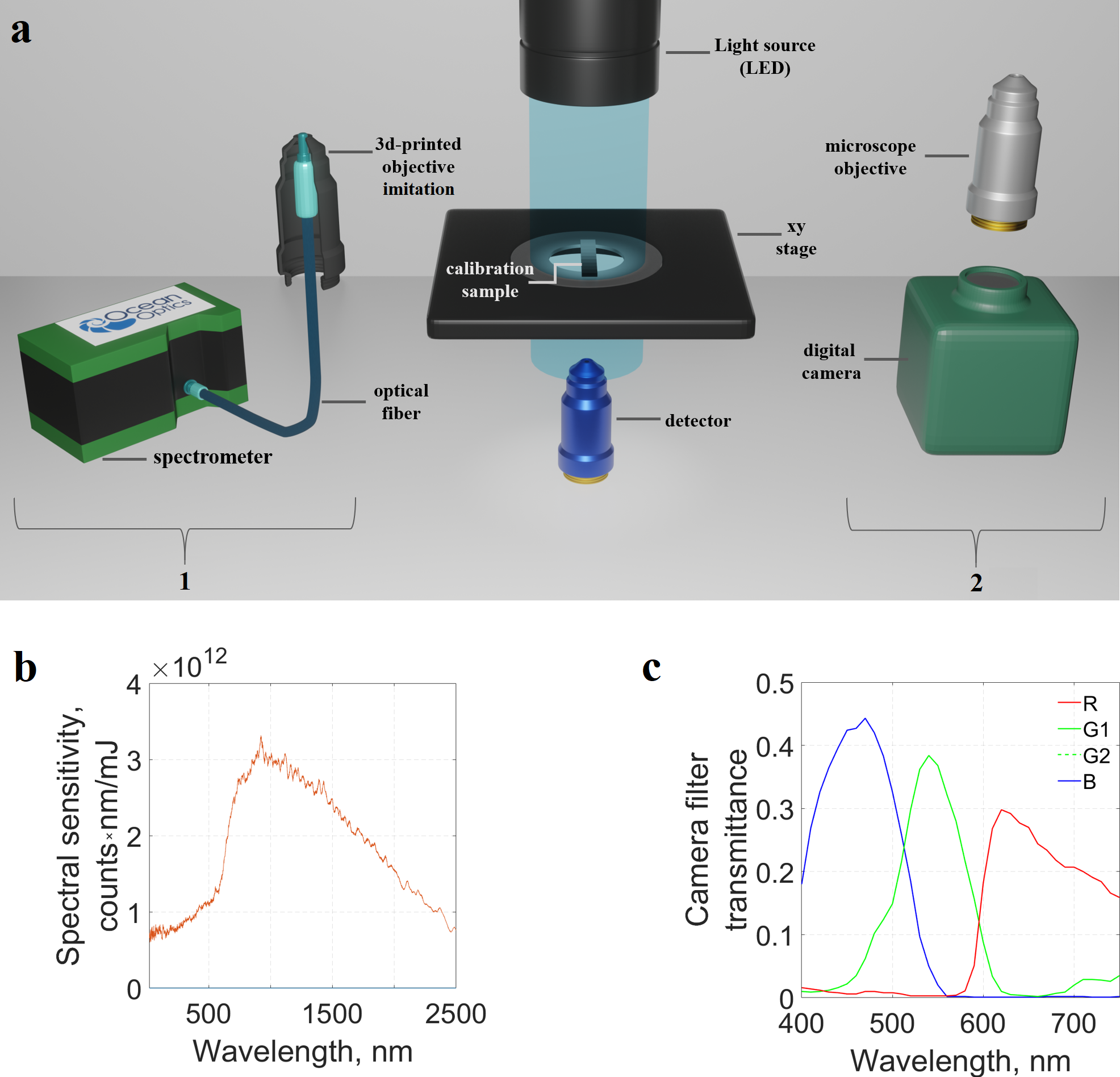}
\caption{(\textbf{a}) The scheme of the measuring apparatus for calibration of the light microscope. 1. A fiber of the optical spectrophotometer placed into a plastic imitation of the microscope objective--arrangement for measurement of the microscope light source spectra after passing through the calibration sample. 2. A real microscope objective and a digital camera as the main part of the arrangement for acquisition of the calibration digital images of the linear step filter. (\textbf{b}) Spectral characteristics of an optical fiber spectrophotometer. The response of the Ocean Optics USB 4000 VIS-NIR-ES spectrometer with a fibre P400-1-UV-VIS after illumination by a NIST calibrated lamp. (\textbf{c}) Spectral characteristics of a color digital microscope camera. The declared spectra of the Bayer mask filters for a Kodak KAI-16000 camera sensor. \label{Fig2}}
\end{figure}

The microscope system that was calibrated is the nanoscope (see Section~\ref{microscope}). The scheme of the measuring apparatus  can be seen in Figure~\ref{Fig2}a. Calibration sample, a Linear Step Filter NDL-10S-4 (Thorlabs, Newton, NJ, USA), is a set of neutral density filters with a discrete range of 10 neutral densities (ND) of optical densities (OD$_{633}$) from 0.1 to 4. The filters themselves are thin layers of gray shades gradients of metallic (NiCrFe), light absorbing, material coated on one side of a 2 mm UV fused silica support. The optical fiber spectrometer used for the calibration of the microscope system was calibrated itself by the procedure described in Appendix \ref{spf_calib} and characterized by its spectral sensitivity (Figure~\ref{Fig2}b). The input data for the calibration were also digitalized microscope camera filters' spectra obtained by the camera producer (Figure~\ref{Fig2}c).  

The camera calibration and image correction was performed in the steps as follows:
\begin{enumerate}
\item Focus position determination:
\begin{enumerate}
\item During  image acquisition, a light is passing through the microscope optical path and sample. The resulting signal was then captured by a camera sensor. Filters of ND 0.1, 0.2, 0.3, 0.4, 0.5, and 0.8 were scanned through their depths with a step of 130.9 nm. In this way, six sets of images were obtained. The results for neutral densities ND <0.1 and ND 0 were acquired for a path of ray with the blank support UV fused silica and for a clear path of ray without any light absorber, respectively;
\item The sets of images for ND <0.1--0.8 were processed by Image Info Extractor Professional software (ICS FFPW) (chosen R\'{e}nyi parameter $\alpha = 2$)~\cite{Ryc16}.
The point information gain entropy (PIE) gives a total change of the image information after iterative removing of one pixel from an individual image.  The highest value of this parameter allows one to pick an in-focus image taken from the center of the coatings (Figure~\ref{Fig3}). This reduced the diffraction-induced aberrations from the boundaries of the calibration sample.
\end{enumerate}

\begin{figure}[H]	 
\includegraphics[width=10 cm]{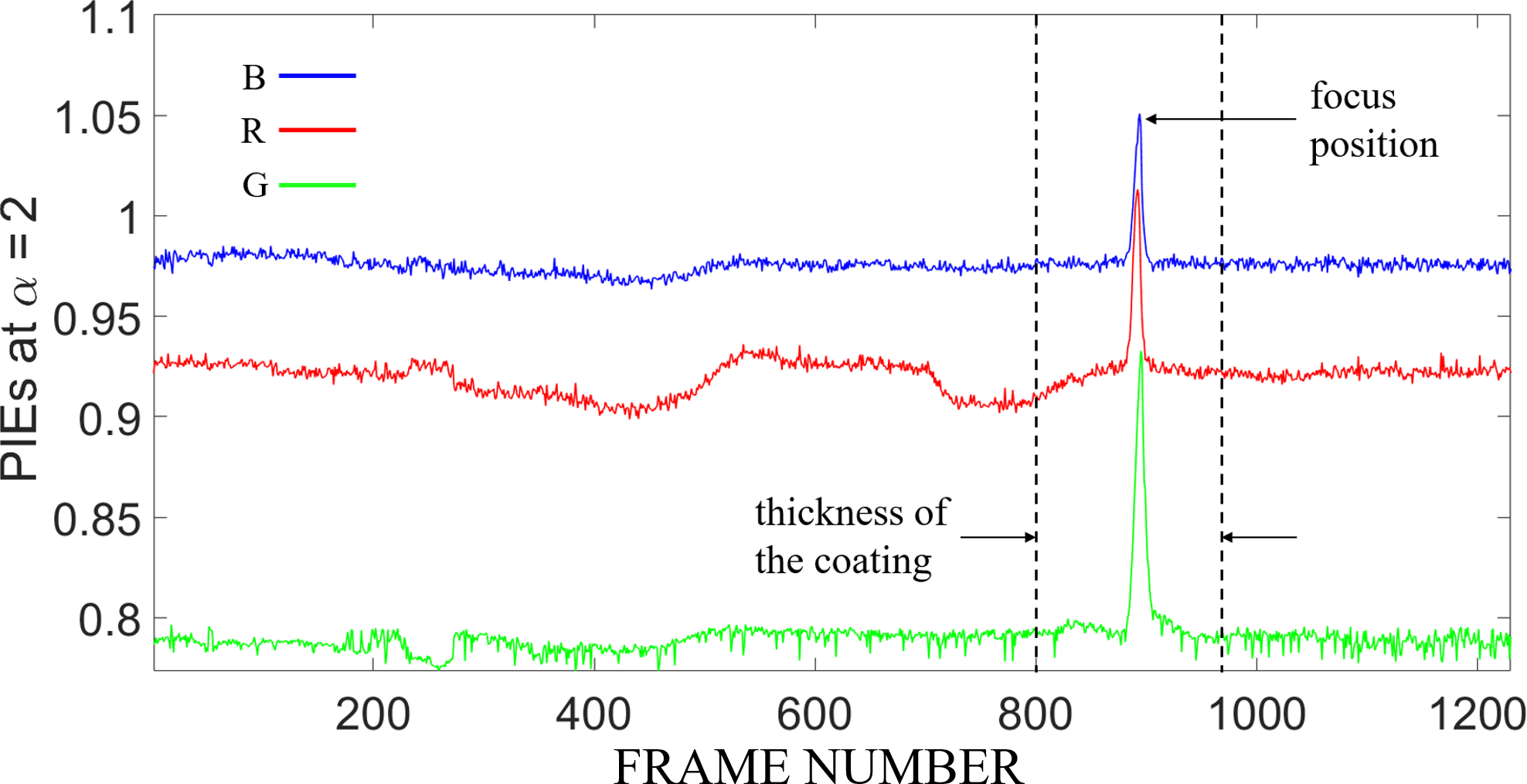}
\caption{The dependence of the PIE at $\alpha$ = 2 on the relative position of the objective along the z-axis for the red (R), green (G), and blue (B) channels of a calibration gray filter (ND 0.3). The z-positions correspond to the frames' numbers. The in-focus image that was selected for the microscope calibration is shown by the arrow. The used microscope objective was a Nikon Plan 40$\times$/0.65, $\infty$/0.17, WD 0.56 at LED current of 4500 mA. \label{Fig3}}
\end{figure}

\item Acquisition of calibration images and transmission spectra of linear step filters:

\begin{enumerate}
\item A set of calibration data (40 images for each neutral density) was taken at the determined focus position (see item 1b). The calibration images were calculated as a pixel-wise mean through the image sets corresponding to each neutral density. In  this way, we reduced a random noise in the measured~data;
\item An optical waveguide P400-1-UV-VIS was housed inside a 3D-printed mechanical adapter at the determined focus position and connected to a spectrometer Ocean Optics USB 4000 VIS-NIR-ES. The spectral responses ND <0.1--0.8 in Figure~\ref{Fig4}a correspond to electroluminescences of relevant calibration sample filters. The spectral response ND 0 reflects electroluminescence captured without any absorber by an optical spectrophotometer during calibration sample illumination.
\end{enumerate} 
\item Calculation of the radiant fluxes reaching each microscope camera pixel after passing the calibration filters.
\begin{enumerate}
\item The raw microscope diode wavelength-dependent signal obtained by a spectrophotometer (Figure~\ref{Fig4}a) was divided by a spectrophotometer sensitivity wavelength-dependent curve (Figure~\ref{Fig2}b). That gave the final calibrated microscope diode signal (Figure~\ref{Fig4}b);
\item After data smoothing and interpolation, the light spectra reaching each pixel of the camera sensor (Figure~\ref{Fig4}c) were obtained as multiplication of the incoming spectra (measured by the calibrated spectrometer; see item 3a and Figure~\ref{Fig4}b) by the respective (red, green, blue) quantum efficiency profile of the Kodak KAI-16000 image sensor (Figure~\ref{Fig2}c);
\item For each neutral density of the filter coating, the radiant flux (Figure~\ref{Fig4}d) reaching each pixel during exposure was calculated as an integral (trapezoidal rule) of the area below the respective incident spectrum.
\end{enumerate}
\item Calibration curves and image correction:
\begin{enumerate}
\item Calibration curves were obtained for each image pixel position and respective extensions of values of camera channel radiant flux (see the range of filter coatings in items 1--2). The calibration curves map the pixel intensities to the real spectra responses (e.g., Figure~\ref{Fig5}). Each pair of consecutive points was interpolated~linearly;
\item In order to extend the range of experimental (useful) intensities, the initial calibration curves for each pixel were extrapolated linearly about 80\% (\mbox{Figure~\ref{Fig5}}, magenta line). Statistics of the pixels' intensity values for each calibration filter was evaluated (Figure~\ref{Fig6}) in order to see the deviations of the sensor pixels' responses from~idealities;
\item  
Hereafter, the calibrated image format (as double precision floating point type) showing the spatial distribution of radiant fluxes was transformed into a png format. The calibration curve associates the value of each calibration intensity with the respective experimental measured photon energy.
\end{enumerate}
\end{enumerate}

\begin{figure}[H]	 
\includegraphics[width=10 cm]{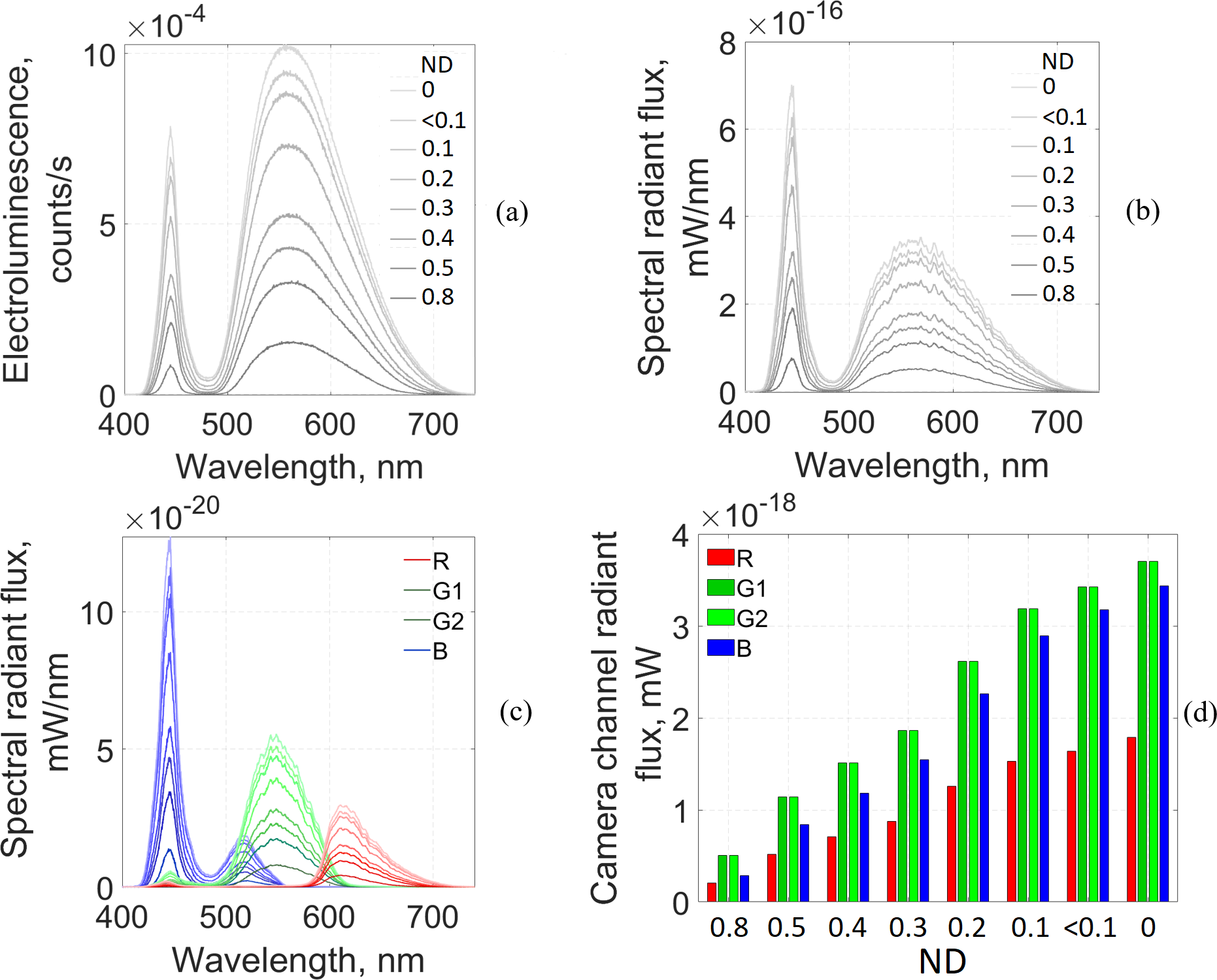}
\caption{(\textbf{a}) The light spectra of gray filter coatings NDL--10S--4 ND 0.1--0.8, the blank silica support (ND < 0.1), and optical path without any absorber (ND 0) measured by a fiber spectrometer. (\textbf{b})~Incoming light spectra corrected to the fiber spectrophotometer spectral sensitivity. (\textbf{c}) The spectra of incoming light captured by the red (R), green (G1, G2) and blue (B) camera pixels. (\textbf{d}) Integral spectral values (radiant fluxes) reaching red, green, and blue camera pixels, respectively, after passing each neutral density ND 0--0.8. \label{Fig4}}
\end{figure}

\subsection{Least Information Loss (LIL) Image Conversion} \label{LIL}

The resulted calibrated images have >8 bpc png format, but nowadays the majority of digital screens can show only 8 bpc formats. For practical purposes, the LIL Convertor software was developed by the ICS FFPW. It converts $>$8 bpc images more informatively and comparably~\cite{Sty16}. This program allows to convert RGB, grayscale, and raw (an appropriate Bayer mask~\cite{Bay76} can be selected) data format to 8 bpc images. Rescaling the intensities in each color (r, g, or b) channel can be applied for image sets of any length (even for one image only) either separately (when the intensity maximum and minimum is found for each color channel separately) or together (one common intensity minimum and maximum, respectively, is found for all color channels). If color channels are normalized separately, more information in the image is preserved. In case of multi-image series, the intensities are rescaled between the maximum and minimum intensities  through the whole series. The empty (unoccupied) levels are always removed in case that they are empty in all images of the image set. The basic LIL algorithm (Algorithm \ref{algorithmA}) is written in Appendix \ref{appendixB}.

The images can be cropped which is useful namely for removing dead pixels' rows and columns. Intensities of, e.g., dead or hot pixels can be eliminated in the image by a thresholding function that, before applying the LIL conversion, replaces the undesirable intensities in an original vice-bit image by a neighbor pixel's active intensity. The program is optimized to utilize multicore CPUs.

\begin{figure}[H] 
\includegraphics[width=10 cm]{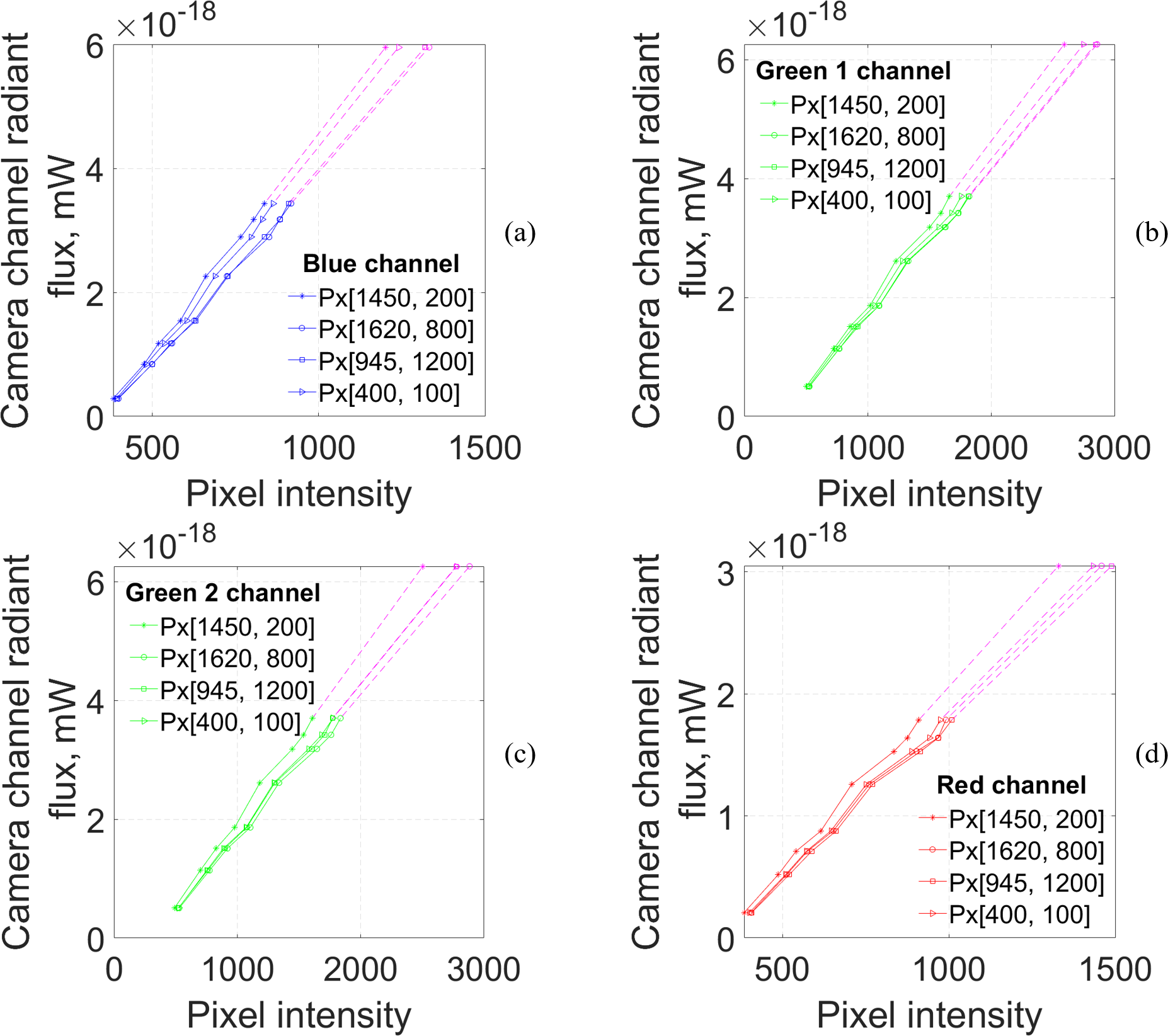}
\caption{Examplary calibration curves for the blue (\textbf{a}), green 1 (\textbf{b}), green 2 (\textbf{c}), and red (\textbf{d}) camera pixels [1450, 200], [1620, 800], [945, 1200], and [400, 100] (counted after image debayerization) for a microscope objective Nikon Plan: 40$\times$/0.65, $\infty$/0.17, WD 0.56 (LED current: 4500 mA, ILCX: $-106$ $\upmu$m, ILCY: $-201$ $\upmu$m).
\label{Fig5}}
\end{figure}

\vspace{-6pt}
\begin{figure}[H]	 
\includegraphics[width=10 cm]{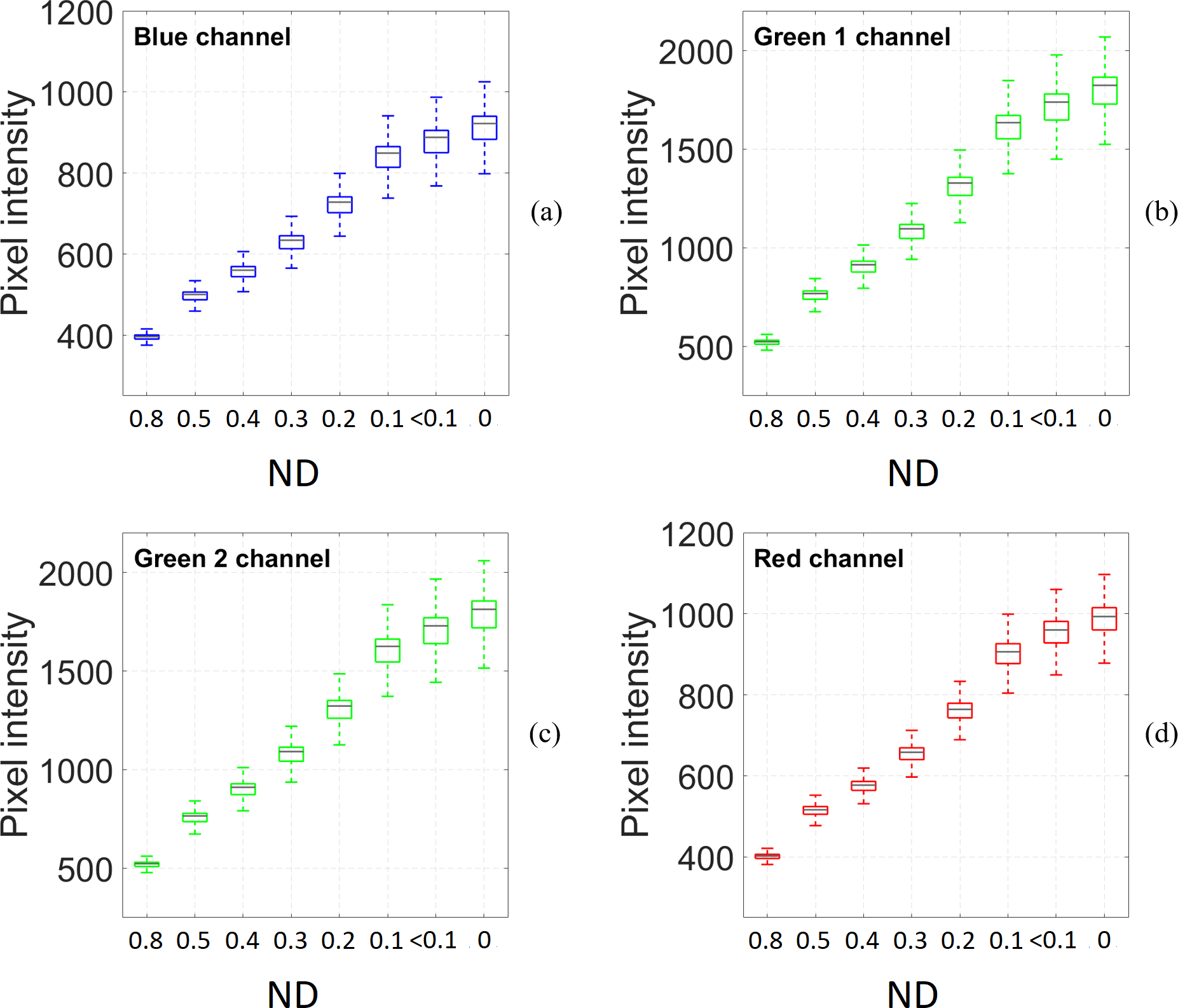}
\caption{Distributions of the calibration image intensities for the blue (\textbf{a}), green 1 (\textbf{b}), green 2 (\textbf{c}), and red (\textbf{d}) camera pixels [1450, 200], [1620, 800], [945, 1200], and [400, 100] (counted after image debayerization) for a microscope objective Nikon Plan: 40$\times$/0.65, $\infty$/0.17, WD 0.56 (LED current: 4500 mA, ILCX: $-106$ $\upmu$m, ILCY: $-201$ $\upmu$m). \label{Fig6}}
\end{figure} 
\section{Results and Discussion}

The nanoscope described in Section~\ref{microscope} was developed mainly for studying unlabeled live and fixed cells to investigate their real structures. Thus, no additional contributions to the object response, such as (imuno)fluorescent dye modifications, influence the detected signal. Investigation of the unmodified samples does not require complicated sample preparation and does not decrease the lifetime of the living cells.

\begin{figure} [H]
\includegraphics[width= 13 cm]{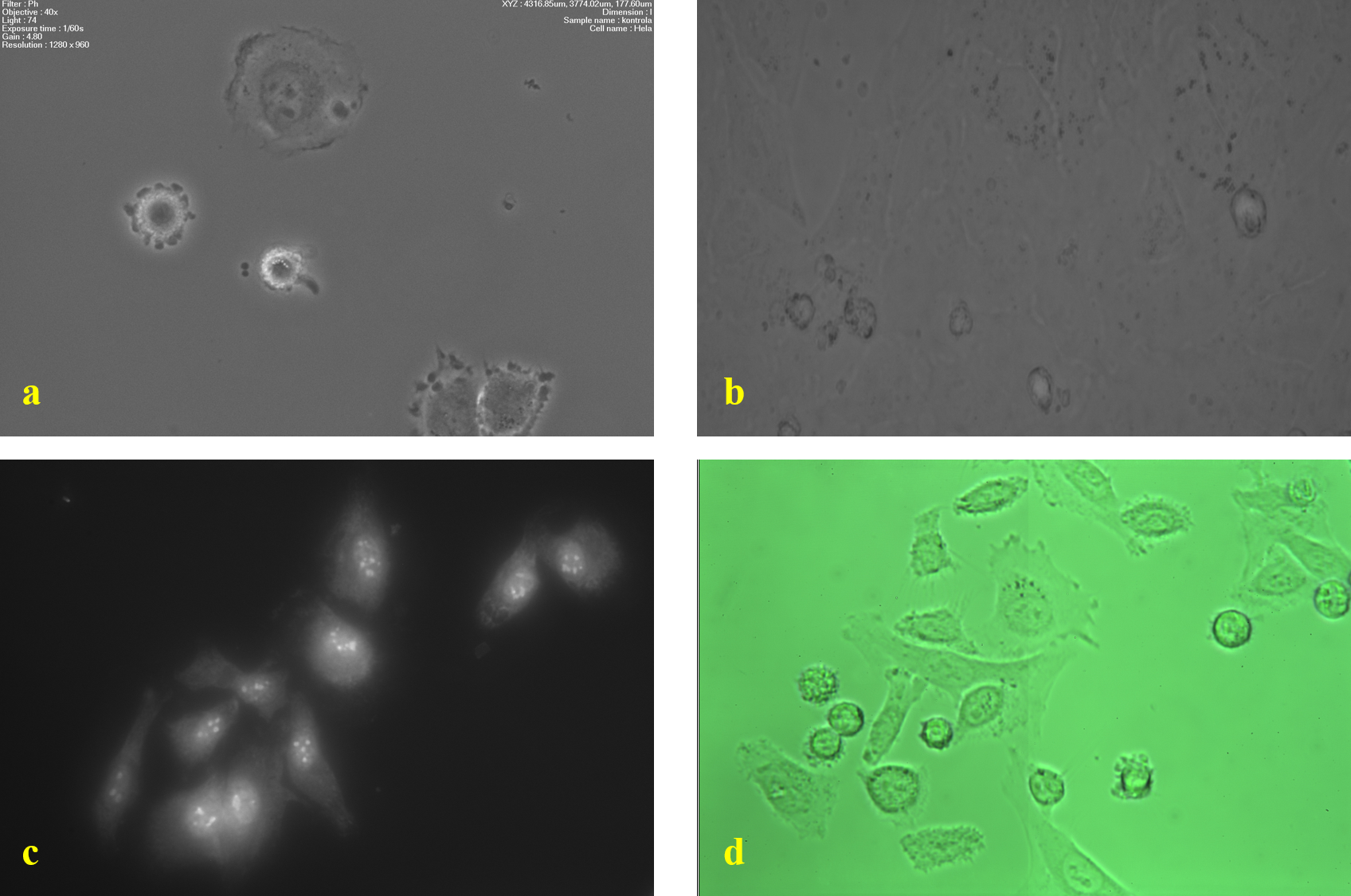}
\caption{Eight-bit-per-channel images of HeLa cells pictured by (\textbf{a}) a phase contrast microscope Nikon BioStation IMQ (40$\times$ objective magnification), (\textbf{b}) a bright-field microscope Olympus IX51 (40$\times$ objective magnification), (\textbf{c}) a fluorescent microscope Nikon Eclipse 80i (40$\times$ objective magnification, stained by fluorescein), and (\textbf{d}) the nanoscope (20$\times$ objective magnification followed by 2$\times$ projective lens magnification) giving raw image data. Panels b and d are visualized by the LIL algorithm. The descriptions of the microscope arrangements are written in Section~\ref{microscope}.} 
\label{Fig7}
\end{figure}

In Figure~\ref{Fig7}, HeLa cells are presented in the phase contrast, bright-field, and   fluorescent modes of a commercial microscope and in video-enhanced high-resolution imaging using the nanoscope at the same magnification. The detail descriptions of the optical paths for the microscopes is written in Section~\ref{microscope}. In the phase contrast imaging (Figure~\ref{Fig7}a), the cell borders are surrounded by halos of light interferences. In the bright-field mode of the commercial microscope (Figure~\ref{Fig7}b), the cell structures are hardly observed, mainly due to a lower intensity of illumination and a larger size of (an object which can be projected on) a camera pixel. Moreover, due to the usage of a grayscale camera, the spectral characteristics of the standard bright-field image can be hardly analyzable. The staining for fluorescence microscopy (Figure~\ref{Fig7}c) visualizes only selected parts of the chemically fixed cells. The cells lost their viability. Moreover, information about the positions, shapes, or behavior of other organelles is also lost. In contrast, the nanoscope (Figure~\ref{Fig7}d) provides primary color images where, namely, all cell borders and condensed chromosomes during mitosis are nicely visible and the contrast can be further intensified by the method of simultaneous optical path and camera chip calibration (Section~\ref{calibration}).

\begin{figure} [H]
\includegraphics[width=10 cm]{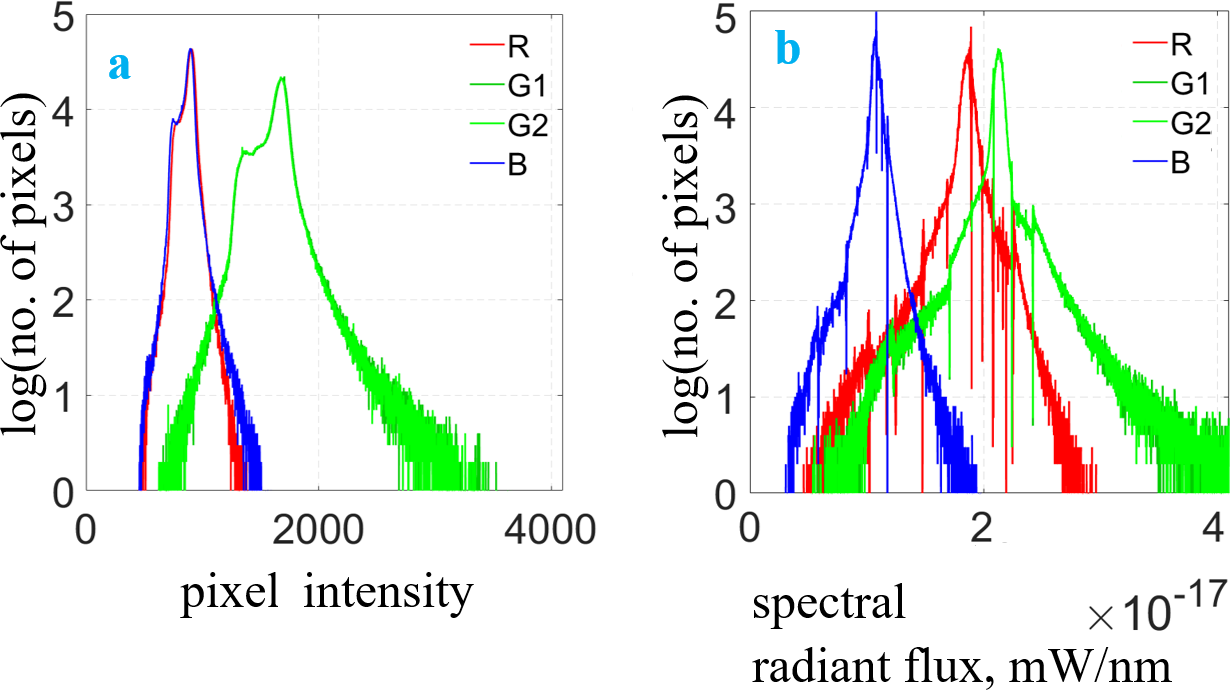}
\caption{Red, green, and blue intensity histograms for raw format of the image in Figure~\ref{Fig9}. (\textbf{a})~Uncorrected raw image; (\textbf{b}) spectrally corrected raw image.}
\label{Fig8}
\end{figure}

The color calibration of a digital camera is a process that has been already known to camera manufacturers for several decades. The fact that the microscope camera is somehow calibrated and, thus, acquired microscopy data are not commonly available in its raw format is unfortunately usually unknown to ordinary users of the digital microscopy. In the proposed calibration method (Section~\ref{calibration}),   the microscope is calibrated, in our case the custom-made nanoscope, as a whole (see Figure~\ref{Fig1})---from the wavelength-dependent transmission and chromatic aberrations of the microscope objective and other optical elements in the microscope optical path up to the microscope camera sensor whose original calibration made by the camera producer was switched off in order to obtain the primary camera signal. The proposed calibration method allows to obtain more realistic spectral properties of a micro-object. The image intensity values in the original image are transformed into values of spectral radiant flux (Figure~\ref{Fig8}). In addition, it is evident (Figure~\ref{Fig8}) that the image histogram after correction is much more structured than that before the correction. Even in absence of the sample, the incoming light is distorted by the optical path and non-idealities of the camera sensor. Moreover, the corrected image shows no signs of optical vignetting (which is undesirable mainly for macroscopic camera imaging purposes; the fields of view of most microscopes are too small to show the edge optical vignetting). Our method helps to mitigate such problems, improving any consecutive method of analysis.

\begin{figure}[H]
 \widefigure
\includegraphics[width= 20 cm]{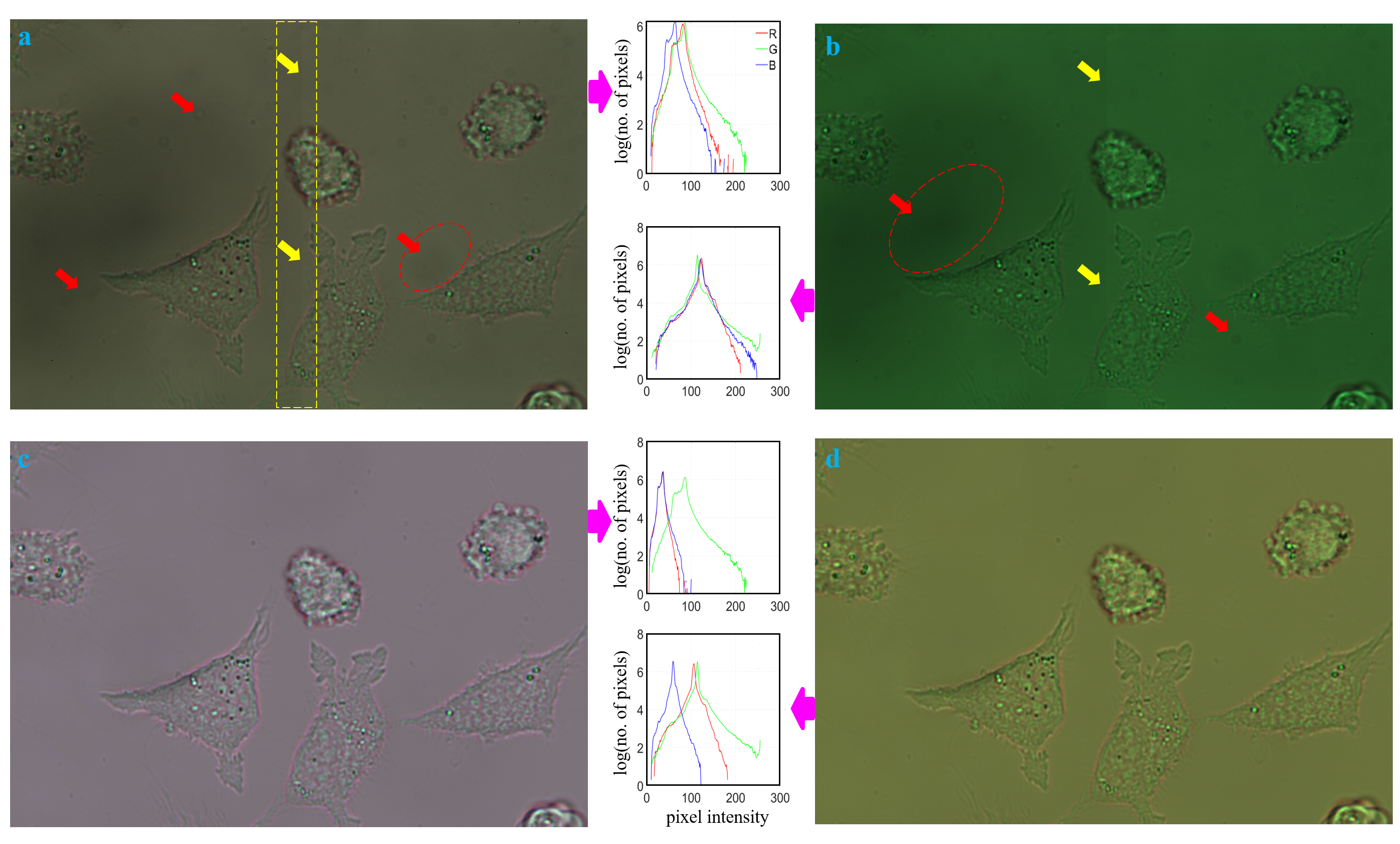}
\caption{The LIL 8 bpc visualization and their corresponding histograms of uncorrected (\textbf{a},\textbf{b}) and spectrally corrected (\textbf{c},\textbf{d}) image of mouse fibroblasts captured by the nanoscope (40$\times$ objective magnification). The LIL transformation was performed for a single image (\textbf{a},\textbf{c}) and through the whole image series (\textbf{b},\textbf{d}). The yellow and red arrows point to the sensor tap imbalance problem and stains of dirty on the microscope optics, respectively. The magenta arrows refer to the relevant histogram of the images (\textbf{a}--\textbf{d}).} 
\label{Fig9}
 \end{figure}

For the correct assessment of experimental results, a proper visualization of the $>$8 bpc images is important. For this, we proposed a LIL algorithm (Section~\ref{LIL}). Figure~\ref{Fig9} shows images of mouse fibroblasts and corresponding image intensity histograms with and without prior calibration converted using the LIL algorithm applied to a single image and a whole time-lapse series. The conditions at which the experiment was performed using the nanoscope are described in Section~\ref{microscope}. There is a visible vertical line in the middle of each uncorrected image. It is caused by sensor taps' imbalance problem, which can be crucial for digital processing algorithms. In addition, in the corrected image, we suppressed artifacts such as grains of dust in the microscope optical path (cf. Figure~\ref{Fig9}a--d) and intensified color contrast between different cellular structures and the background. All images in \mbox{Figure~\ref{Fig9}} were obtained by the separate normalization of color channels and thus there are no natural colors. Even though the information is necessarily lost in the transformation from $>$8 bpc into 8 bpc images, the LIL transformation of an individual image preserves the highest proportion of it and the transformation of a whole series preserves most of the information in the series and allows a mutual visual comparison of its images.

\startlandscape
\finishlandscape

\begin{figure} [H]
\includegraphics[width=13 cm]{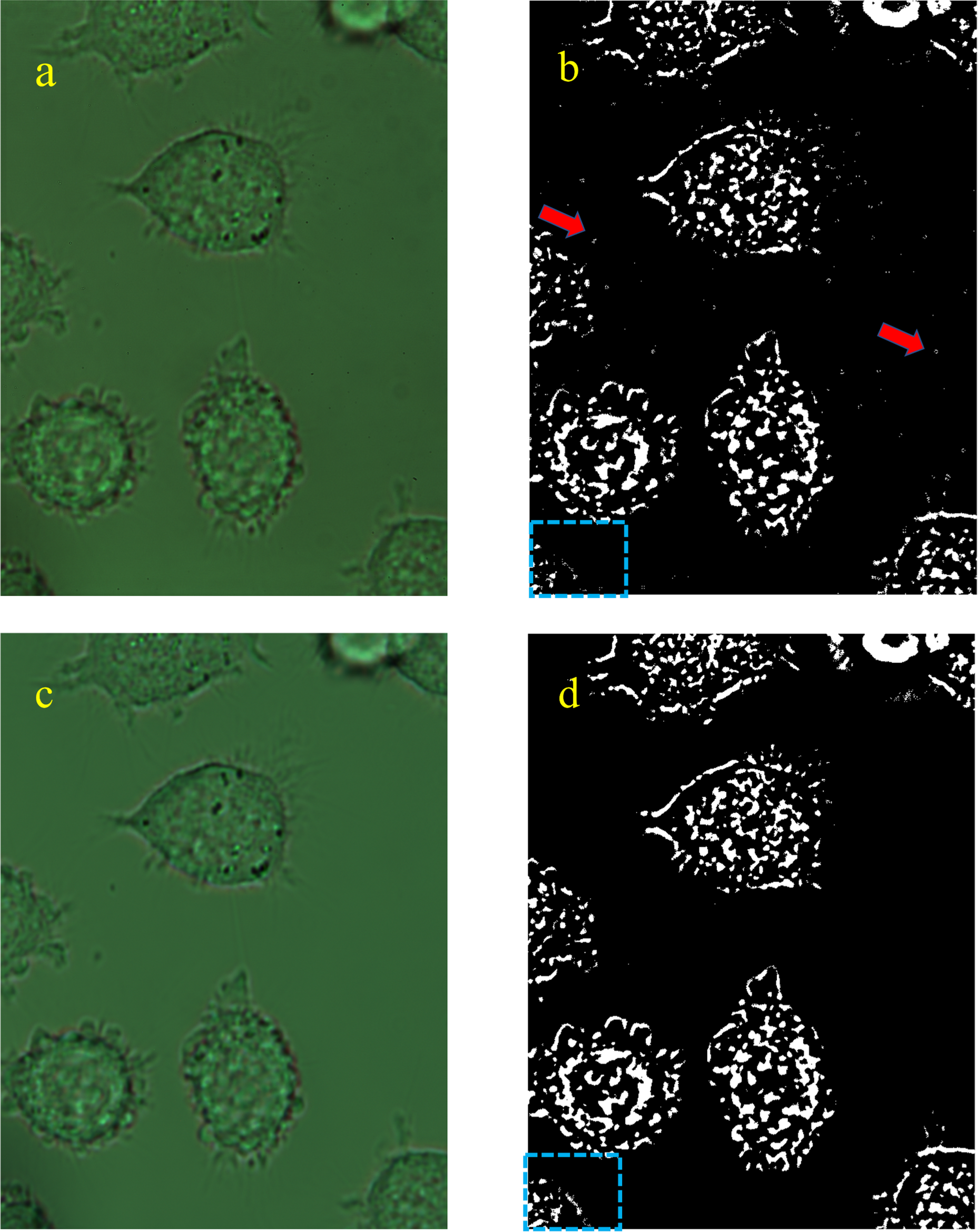}
\caption{Conversions of a mouse fibroblasts' image using simple Matlab functions. (\textbf{a}) 8-bit representation of a 12-bit-per-channel original uncorrected rgb image, (\textbf{b}) binarized image, (\textbf{a},\textbf{c}) 8-bit representation of a corrected rgb image, (\textbf{d}) binarized image (\textbf{c}).} 
\label{Fig10}
\end{figure}

As an example of how the described method affects the subsequent image processing, we applied a standard Matlab binarization function (after color-to-grayscale image transformation) to the corrected and uncorrected images (Figure~\ref{Fig10}). Figure~\ref{Fig10} was acquired on the same Nikon 40$\times$ objective as Figure~\ref{Fig9}. The uncorrected rgb and binarized images (Figure~\ref{Fig10}a,b) have background non-uniformities (scratches and dust in the microscope optical path shown by arrows) and also presents  less compact borders of pseudopodia (shown in the square) when compared with the corrected images (Figure~\ref{Fig10}c,d).

Conceptually, the most similar to our approach of microscope calibration are procedures described in~\cite{Thi10,You08,Gru08}.

Thigpen et al. described a photometric calibration which is suitable for a fluorescence microscope or a transmitted light microscope~\cite{Thi10}. The method works wavelength- and exposure-wise, establishing white balance in the image background. The method was demonstrated on a microscope equipped with a grating-based spectral light source and a grayscale digital camera. Absorption spectra originated from various filters placed in the computer controllable emission-side filter wheel of the microscope. Calculating the average optical density (i.e., average intensity) of the background surrounding a sample for an image at a fixed wavelength allowed the researchers to approximate the incident light to the filter. Performing this calculation for all wavelengths yielded an intensity profile of the incident light which was flat through all wavelengths. After selection of a reference image (at a wavelength of 550 nm and reference camera exposure), the similarity measure between the reference image and the testing image was calculated for an image captured at each combination of wavelength and camera exposure. Then, the exposure for the image, which was the most similar to the reference, was selected as an exposure at the relevant wavelength.

Since this calibration method~\cite{Thi10} is wavelength-wise, it allows the segmentation of objects of a particular wavelength in a microscopy image like our spectral method (following our calibration approach,~\cite{Lon20}). Unlike our method, the calibration filters are not placed in the focal plane of the imaging system. Moreover, this method~\cite{Thi10} can show a drift in the monochromator's precision.

Young et al.~\cite{You08} proposed a method for photocalibration of a fluorescence microscope. The authors employed a set of six spectrally different LEDs and varied the current through the LEDs in a controlled way. A controlled amount of light emitted by the LEDs and transmitted by the microscope optical path (including/excluding combination of filters of different ND) was measured by a grayscale image sensor. After that, the digitized intensity was explained as a function of the LED current. The nonlinearity in the dependency ``image intensity vs. LED current'' was suppressed by the calibration of the excitation light by a photodiode using a Spectralon reflector. The photodiode has a dynamic range of 10$^5$:1 and its integration time and gain were computer controllable.

Compared with this method~\cite{You08}, our approach is also suitable for not only bright-field microscopy but also fluorescence microscopy. However, our method is pixel-wise. Moreover, we utilize spectral properties of color camera chip filters which allows us to obtain complete information on spectral properties of imaged objects in each individual image. In contrast the approach described in~\cite{You08} requires to capture a set of images in each time step. By our method, we  can thus obtain spectral information also about objects that vary quickly over time. In addition, our method allows us calibration of the optical path in light energies. Such a calibration by Young's approach is substantially technically more complicated. For that, the authors proposed a special portable instrument of the size of a microscope slide that uses LEDs at different wavelengths and with controllable amount of current going through in order to produce calibrated amounts of light~\cite{You06}. Unfortunately, for work~\cite{You08}, we have not found any image outputs which would allow us qualitative and quantitative comparisons with our method.

The authors of~\cite{Gru08} described a calibration of light fluxes for a wide-field fluorescence microscope using a thermocouple power meter, after passing the light through an objective with an openable iris. The method allowed the researchers to precisely estimate excitation intensities in the object plane using commercially available opto-mechanical components, compare experimental set-ups, and highlight details in an image. The whole calibration process for one objective and six fluorescence filter cubes lasted for 75 min.
This method seems to be the most similar to ours. The main difference is in the usage of the thermocouple power meter for measurement of doses of the incident light instead of a fiber optical spectrophotometer. Using this procedure~\cite{Gru08}, one cannot obtain any spectrally resolved image information.

\section{Conclusions} \label{conclusions}

Biological experiments are hardly reproducible and repeatable but also often sensitive to the technical provision. The approach presented in this paper allows, without any extensive technical treatment, one to reduce, for one thing, the impact of optical path inhomogeneities and, for another thing, defects of a camera sensor on color properties of images in a bright-field light transmission microscopy experiment. Thus, it allows one to maximize the yield of information obtained in the technically simplest possible experiment. The calibration procedure proposed here (if needed, followed by a correct image intensity compression) is promising for more adequate and precise transfer of color information to the computer and researchers. This experimental technology allows observation of living cells in dynamics and makes the following microscopy image processing (especially, cells' features and edge detection) easier.

The information in the uncalibrated image is obviously dominated by the sensor and light path inhomogeneities and for proper evaluation only calibrated images may be used for assessment. This is particularly important for the eventual automatic machine analysis which will be unavoidable in case of a large field of view containing many individual objects that cannot be analyzed manually.

This method    makes  edge detection (extraction of the edge features) easier, which is important in image processing. In other words, the calibration method allows a better grouping of the intensities of similar spectral properties in order to separate background from foreground in a bright-field transmission image.

This calibration approach as such is independent of the light microscope construction, from all its optical components to a digital camera type (monochromatic vs. color, of different bit depths). After slight differences in the procedure, this calibration method can be also suitable for contrast microscopy or fluorescent microscopy. The only condition of a successful application of the method is a microscope camera giving a primary (raw) signal which is not distorted and modified by either electronics or software.

Nevertheless, the described image calibration and correction method in bright field is not applicable only for observation of a live cell but also for analysis of other transparent materials such as, e.g., polymer scaffolds or nanoprinted polymer structures. After some slight changes in the experimental protocol and system arrangement, the method can also be usable in light reflection or fluorescence microscopy.

It can be concluded that, compared with the microscope calibration methods published previously, our approach to calibration is the most versatile and also enables   further advanced data processing including spectral image data analysis.

\vspace{6pt}



\authorcontributions{Conceptualization, G.P., D.\v{S}. and R.R.; methodology, D.\v{S}. and R.R.; software, P.S. and K.L.; validation, G.P., D.\v{S}. and R.R.; formal analysis, G.P., D.\v{S}., and R.R.; investigation, G.P., D.\v{S}., R.R. and J.V.; resources, D.\v{S}.; data curation, G.P.; writing---original draft preparation, G.P.; writing---review and editing, R.R.; visualization, G.P.; supervision, D.\v{S}. and R.R.; funding acquisition, D.\v{S}. All authors have read and agreed to the published version of the manuscript.}

 
\funding{This work was supported by the Ministry of Education, Youth and Sports of the Czech Republic--projects CENAKVA (LM2018099) and from the European Regional Development Fund in frame of the project Image Headstart (ATCZ215) in the Interreg V-A Austria--Czech Republic programme. The work was further financed by the project GAJU 017/2016/Z. One of the authors (J.V.) acknowledges funding from the Charles University centre UNCE/SCI/010.}
\dataavailability{The {supplementary} data, namely Figures~\ref{Fig2}, \ref{Fig4}, \ref{Fig5}, and \ref{Fig9} from other nanoscope set-ups, calibration sets, and Matlab scripts, are available in~\highlighting{\cite{dryad}}.}

\institutionalreview{Not applicable.
}
\informedconsent{Not applicable.

}

\acknowledgments{The authors would like to thank Petr Mach\'{a}\v{c}ek (ImageCode, Brloh, CZ), Vladyslav Bozhynov, Ali Ghaznavi, \v{S}\'{a}rka Beranov\'{a} and Pavl\'{i}na Tl\'{a}skalov\'{a} (all from ICS USB) for their support of this paper.}

\conflictsofinterest{The authors declare no conflict of interest. The funders had no role in the design of the study; in the collection, analyses, or interpretation of data; in the writing of the manuscript, or in the decision to publish the~results.}
\newpage
\abbreviations{Abbreviations}{The following abbreviations are used in this manuscript:\\

\noindent 
\begin{tabular}{@{}ll}
bpc & bits per channel\\
PBS & fetal bovine serum\\
FFPW & Faculty of Fisheries and Protection of Waters\\
ICS & Institute of Complex Systems\\
ILCX & x-position read out by a microscope incremental linear sensor\\
ILCY & y-position read out by a microscope incremental linear sensor\\
LED & light-emitting diode\\
LIL & Least Information Loss algorithm\\
LWD & microscope objective low working distance\\
N.A. & Numerical Aperture\\
NAMC & Numerical Aperture Modulation Contrast iris\\
NIST & National Institute of Standards and Technology\\
ND & Neutral Density\\
OD$_{633}$ & Optical Density at 633 nm, OD$_{633} = -\log$ T$_{633}$, where T$_{633}$ is a light transmission\\
 & at 633 nm\\
PBS & phosphate buffer saline\\
PIE & Point Information Gain Entropy\\
PIG & Point Information Gain\\
PNG & portable network graphics\\
rH & relative humidity\\
rpm & rotations per minute\\
UV & ultraviolet\\
VIS-NIR-ES & visible and near-infrared with enhanced sensitivity\\
WD & microscope objective working distance\\
\end{tabular}}

\appendixstart
\appendix
\appendixtitles{yes}
\section{Optical Fiber Spectrophotometer Calibration} \label{spf_calib}

\begin{figure}[H] 
\includegraphics[width=9cm]{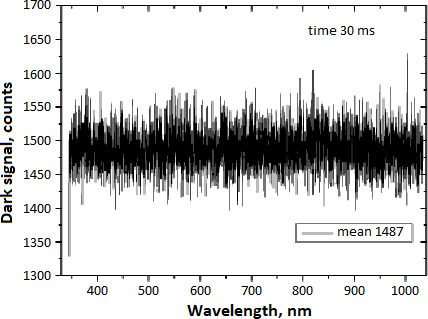}
\caption{Dark signal of the spectrometer Ocean Optics USB4000-VIS-NIR-ES. \label{FigA1}}
\end{figure}

A new tungsten-halogen calibration lamp (a NIST standard of spectral irradiance) Oriel Mod. 63358, S/N:30024 with a power supply Oriel OPS-Q250 (stabilized at 6.5 A for more than 1 h) was used as a calibration standard. The calibrated device was a spectrometer Ocean Optics USB4000-VIS-NIR-ES (S/N USB4U06341) with a fibre P400-1-UV-VIS (1 m long, 0.4 mm core, SMA connectors). The distance between the lamp and the entrance of the calibrated system (fibre input) was 50 cm. For the measurements, the parameters of spectrophotometer control software Ocean View was set up as follows: time---30 ms, electric dark---OFF, average number of scans---3. The signal was above half of the full range (which is a 16-bit number). The dark signal was measured separately (Figure~\ref{FigA1}).

From the experimental data, the wavelength-dependent net signal rate [count$\cdot$s$^{-1}$] was calculated as $\mbox{NSR}(\lambda) = \frac{\mbox{light signal}(\lambda)-\mbox{dark signal}(\lambda)}{0.03}$. This was divided by the lamp spectral irradiance (the calibration provided by the NIST for 50 cm distance gives in the visible spectral region 400–800 nm a monotonously increasing function from about 0.43 to 7.56 mW$\cdot$cm$^{-2}\cdot$nm$^{-1}$) multiplied by a fibre core area (1.25 $\times$ 10$^{-3}$ cm$^2$). This gives an input spectral power [mW$\cdot$nm$^{-1}$]. The resulted value is the system response (or sensitivity, Figure~\ref{Fig2}b) in [count$\cdot$nm$\cdot$mW$^{-1}\cdot$s$^{-1}$] = [count$\cdot$nm$\cdot$mJ$^{-1}$]. The reflection of light on the fibre--air interface is not corrected for as it is part of the system response.

Several measurements with bandpass filters were performed in order to obtain information on stray light effects of the spectrophotometer (they are relatively low) and the final response is corrected for it.

\appendix
\section{Least Information Loss Matlab Algorithm and Pseudocode\label{appendixB}}\vspace{-6pt}
\setcounter{algocf}{0}
\renewcommand\thealgocf{A\arabic{algocf}}							
\setlength{\algotitleheightrule}{0.5pt} 

\begin{algorithm}[!h] 
 
\caption{The Least Information Loss algorithm for a grayscale image.} \label{algorithmA}
\IncMargin{1em} 
\LinesNumbered
\KwIn{a set of \textbf{N} input $n$-bit grayscale images, where $n>8$;
}
\KwOut{\textbf{I8} as a Least Information Loss transformed image of the image In}

\BlankLine
\BlankLine

\textbf{H} = zeros(2$^n$, 1); \qquad\quad\emph{\% create an empty (zero) vector of the length n}

\BlankLine
\BlankLine

\For{$i = 1$ \KwTo N}{
\textbf{In} = readIm($i$); \qquad\emph{\% read an i-th n-bit image}\\
\textbf{h} = createHist(\textbf{In}); \emph{\quad\% create the image n-bit histogram} \textbf{h};\\
\textbf{H} = \textbf{H} + \textbf{h}; \qquad\qquad\emph{\% create the series histogram} \textbf{H};
}

\BlankLine
\BlankLine

\textbf{OccL} = findOccLev(\textbf{H});\\
\qquad\emph{\% write positions of all occupied intensity levels in histogram} \textbf{H}; \emph{omit positions of\\ \qquad \quad the empty histogram bins};\\

\BlankLine

\textbf{NoOfOccL} = length(\textbf{OccL});\\
\qquad\emph{\% calculate the number of the occupied intensity levels} \textbf{OccL} \emph{in the series\\ \qquad \quad histogram} \textbf{H};\\

\BlankLine

\textbf{sf} = ({\textbf{NoOfOccL}-1}/255);\\ 
\qquad\emph{\% compute a scale factor} \textbf{sf} \emph{which says how many occupied levels of an image} \textbf{In} \\ \qquad \quad \emph{will fall in a single, 8-bit-image, intensity; use the floor function};

\BlankLine
\BlankLine

\textbf{In} = readIm(j); \qquad\emph{\% read an j-th n-bit image, where j $\in [1, N]$}\\
\textbf{I8} = \textbf{In} .* 0; \qquad\emph{\% create a zero matrix of the size of image} \textbf{In}\\

\BlankLine
\BlankLine

\textbf{intst} = 0;\\
\BlankLine
\For{i = sf \KwTo NoOfOccL \KwBy sf}{
\textbf{I8}(\textbf{In} == \textbf{OccL}(i)-1) $\gets$ \textbf{intst}/\textbf{sf};\\
\qquad \emph{\% write intensity} \textbf{intst} $\in [1, 255]$ \emph{at the position of the image} \textbf{I8} \emph{which\\ \qquad \quad corresponds to the positions in image} \textbf{In} \emph{with intensities that are higher than\\ \qquad \quad intensity at the position i in the vector} \textbf{OccL}\\
\textbf{intst} = \textbf{intst} + 1;
}

\BlankLine
\BlankLine
 
\end{algorithm}

\end{paracol}
\reftitle{References}

%


\end{document}